\documentclass[sn-apa]{sn-jnl}

\usepackage[utf8]{inputenc}
\usepackage[english]{babel}
\usepackage{csquotes}
\usepackage{graphicx}
\usepackage{bm}
\usepackage{amsthm}
\usepackage{amsmath}
\usepackage{multirow}
\usepackage{makecell}
\usepackage[caption=false]{subfig}
\usepackage{booktabs}
\usepackage{soul}
\usepackage{tabularx}
\usepackage{arydshln}
\usepackage{pdflscape}
\usepackage{adjustbox}
\usepackage{float}
\usepackage[scientific-notation=true]{siunitx}

\usepackage{xurl} 
\usepackage{hyperref}
\hypersetup{breaklinks=true} 

\normalbaroutside
\usepackage{todonotes}

\jyear{2023}%

\theoremstyle{thmstyleone}%
\theoremstyle{thmstyletwo}%

\theoremstyle{thmstylethree}%
\newtheorem{definition}{Definition}%

\begin{document}

\title[Exploring Investor Behavior in Bitcoin: A Study of the Disposition Effect]{Exploring Investor Behavior in Bitcoin: A Study of the Disposition Effect}


\author*[1]{\fnm{J{\"u}rgen E.} \sur{Schatzmann}}\email{juergen.schatzmann@research.usc.edu.au}

\author[2]{\fnm{Bernhard} \sur{Haslhofer}}\email{haslhofer@csh.ac.at}
\equalcont{These authors contributed equally to this work.}


\affil*[1]{\orgdiv{School of Business and Creative Industries}, \orgname{University of the Sunshine Coast}, \orgaddress{\street{90 Sippy Downs Drive}, \city{Sippy Downs}, \postcode{4556}, \state{Queensland}, \country{Australia}}}

\affil[2]{\orgname{Complexity Science Hub Vienna}, \orgaddress{\street{Josefstädter Straße 39}, \city{Vienna}, \postcode{1080}, \state{Vienna}, \country{Austria}}}

\abstract{
Investors commonly exhibit the disposition effect --- the irrational tendency to sell their winning investments and hold onto their losing ones.
While this phenomenon has been observed in many traditional markets, it remains unclear whether it also applies to atypical markets like cryptoassets.
This paper investigates the prevalence of the disposition effect in Bitcoin by using transactions targeting cryptoasset exchanges as proxies for selling transactions.
Our findings suggest that investors in Bitcoin were indeed subject to the disposition effect, with varying intensity. They also show that the disposition effect was not consistently present throughout the observation period. Its prevalence was more evident from the boom and bust year 2017 onwards, as confirmed by various technical indicators.
Our study suggests irrational investor behavior is also present in atypical markets like Bitcoin.
}




\maketitle

\section{Introduction}
\label{sec:introduction}

The \emph{disposition effect} is a well-known phenomenon in behavioral finance and is observed when investors tend to sell their winning investments and hold onto their losers~\citep{Statman1985}. Its presence has been shown in a wide range of established traditional markets such as the stock market~\citep{Grinblatt2001}, treasury bonds~\citep{Coval2005} or the real estate market~\citep{Genesove2001,Shapira2001} to name a few.

However, to the best of our knowledge, there is no comparable empirical evidence for cryptoasset markets like Bitcoin, which can be considered as being \emph{atypical} compared to established markets: they lack a single central authority, they are still largely unregulated, and they show unusually high volatility~\citep{Doguet2013,EbaVc2016,Nakamoto2008,Chuen2015a}. Hence, we formulate the central research question to be answered in this paper as follows:

\begin{quote}
\begin{center}
\emph{
	Do investors in atypical cryptoasset markets tend to sell their winning positions more frequently than their losing positions?
}
\end{center}
\end{quote}

Recent related research by~\cite{Baur2018}, who investigated the effect of \emph{fear of missing out} in several cryptoasset markets, already suggests that there might be a link between volatility and the disposition effect but does not provide empirical evidence for its existence.

Therefore, we approach this research question by examining and validating this economic pattern at the transaction level~\citep{DiFrancescoMaesa2017,Ranshous2017,Ober2013}. Doing this is possible in cryptoassets since transaction data is openly available, and we can also use several methods~\citep{Meiklejohn2013} and tools~\citep{Haslhofer2016,Kalodner2017} that enable us to investigate the transaction behavior of economic actors. We are particularly interested in interactions with cryptoasset exchanges. They play a central role in trading cryptoassets because they allow investors to buy and sell cryptoassets.

\paragraph{Contribution}

In this paper, we expand on existing research and empirically investigate the prevalence of the disposition effect in Bitcoin. We formulate our overarching hypothesis as follows:



\begin{quote}
	\begin{center}
	\textbf{$H_1$}: The frequency of Bitcoin sells is \textbf{higher} in \textbf{positive} market conditions than in \textbf{negative} market conditions.
	\end{center}
\end{quote}

The corresponding null hypothesis is defined as \emph{$H_0$: There is no significant difference in the frequency of Bitcoin sells between positive and negative market conditions.}

To test this hypothesis, in Section~\ref{sec:DataAndMethod}, we first develop a method that allows us to measure \emph{Gains Realized (GR)} compared to \emph{Losses Realized (LR)} when Bitcoins are sold at some cryptoasset exchange. Our method is inspired by~\cite{Odean1998}, who measured \emph{Proportion of Gains Realized (PGR)} and \emph{Proportion of Losses Realized (PLR)} in individual investment portfolios to test for the disposition effect. As our research investigates a single highly volatile cryptoasset instead of a portfolio of relatively stable assets in a traditional market, our metrics are non-proportional and computed based on hourly cryptoasset exchange rates.

Next, in Section~\ref{sec:results}, we apply this method on the entire blockchain from its inception until November 24th, 2021. We also compute well-established technical indicators such as the \emph{Relative Strength Indicator (RSI)} or \emph{Moving Average Convergence Divergence (MACD)} to allow for comparison. Our empirical result supports the existence of the disposition effect with varying intensity for the Odean metrics and most of the other technical indicators.

Our results provide evidence that, during multiple time frames within our observation period from Bitcoin's inception until November 2021, cryptoasset traders exhibited irrational selling behavior similar to that observed in established markets, suggesting the presence of the disposition effect in the cryptoasset market during these specific time frames. The boom and bust year 2017 was a pivotal point in the investors' trading behavior. Disposition biased trading significantly increased, attributed to increased media coverage and publicity, leading to a significant inflow of new investors and an uplift in the number of sell transactions in the market.

For reproducibility, we make our dataset and our implementation available at~\url{https://github.com/jschatzmann/CryptoDisposition}.

\section{Background}
\label{sec:background}
 
We will now explain the principle of the disposition effect, which is a well-known phenomenon in behavioral finance, and briefly consider how it was tested in previous work. In addition, we will give a brief introduction to cryptoassets and explain how our measurements expand the existing body of knowledge.

\subsection{Disposition Effect}

Behavioral finance provides the theoretical foundation for this research as it tries to explain the inefficiencies assumed prevalent in the market~\citep{Bruce2010}. Those inefficiencies are described as under- and overreaction to market news and are rooted in the limited attention of (uneducated) investors active in the market. Overreaction occurs when the market reacts too strong or too long to the news, and therefore the adjustment in the opposite direction is required. This phenomenon is also seen in customers' irrational purchasing habits~\citep{Tang2013} or when investors underreact to news that generate a predictable price drift~\citep{Frazzini2004}.

The disposition effect originates from prospect theory~\citep{Kahneman1979} and can be described as \emph{``[...] consistent with the predictions of prospect theory. There is compelling evidence that investors tend to sell their winning investments and to hold onto their losers.''}~\citep{Barber2005}. In effect, traders trade too much due to overconfidence.

\cite{Statman1985} coined the term \emph{``disposition effect - the predisposition to get-evenitis''}~\citep{Shefrin2007} and set the aspects of mental accounting, regret or loss aversion, and self-control into a broader theoretical framework. They describe that investors are keeping separate mental investment accounts. The value function of prospect theory~\citep{Kahneman1979} stipulates that people are generally more loss-averse, which leads towards the disposition effect when applied on the stock market. 

Most relevant for this research is the study conducted by \cite{Odean1998}, who investigated 10,000 randomly selected traders of a trading platform and calculated the proportion of gains compared to the proportion of losses of individual accounts in order to test for the disposition effect. Apart from confirming the presence of the disposition effect, Odean's research provides the theoretical and methodological foundation for the current study.

Assuming that traditional market rules and trading patterns also apply to new and super-volatile asset classes like Bitcoin, it is reasonable to hypothesize that this also applies to well-known bevioral biases, like the disposition effect. To test this hypothesis, we use established measurement instruments, such as Technical Analysis, to objectively quantify market conditions. Additionally, we use transactions targeting cryptoasset exchanges as proxies for selling transactions. By combining these methods, we can investigate the prevalence of the disposition effect under specific market conditions. The results will complement existing literature and contribute to a better understanding of the impact of behavioral biases in the cryptoasset market.

\subsection{Technical Analysis and Indicators}
\label{sec:TechIndicators}

Next to the average price indicator inspired by \cite{Odean1998}, we also apply well-known technical analysis (TA) methods and related indicators, which are applied in established traditional markets. This approach is assumed reasonable as previous research categorizes Bitcoin as a speculative asset rather than as any means of payment
~\citep{Glaser2014,Baur2015,Chu2015,Mullan2014,VanWijk2013,Tu2015,Chuen2015a}.

Technical analysis and the underlying technical indicators are used to investigate and examine a stock market from a purely statistical point of view and therefore play a fundamental and similarly valuable role in the daily work of financial analysts~\citep{Kirkpatrick2007}. \cite{Murphy1999} explains the principles of TA and defines it as follows: \textit{``Technical analysis is the study of market action, primarily through the use of charts, for the purpose of forecasting future price trends.''}

The role of TA is to help analysts to determine when various markets have turned in a primary way. The aim is to identify trends at the earliest stage to maintain the investment posture until indicators determine that the trend has reversed~\citep{Pring2014}. Key assumptions in TA are built on the principles presented in the literature as follows~\citep{Caginalp2003}:

\begin{enumerate}
    \item \textbf{Price trends tend to persist}, essentially capturing the momentum concept, stating the supply/demand ratio is slowly varying.
    \item \textbf{Market action is repetitive}, conceptualizing the fact of recurring patterns in price charts that are evolving due to the consequence of investors' reactions.
\end{enumerate}

Competitive investors continuously strive to ``beat the market''. Investors can always find rewards in (financial) markets as the markets themselves are inefficient but to an efficient extent: competitive investors push the market toward efficiency but without ever getting there~\citep{Daly2017}. Bitcoin, like stock markets, also follows the efficient market hypothesis, and prices react immediately to publicly announced information~\citep{Bartos2015,Clark2016}. Hence we believe Bitcoin is very attractive to speculative and risk-seeking investors using TA with related buy and sell rules. We provide a list of the selected technical indicators and the related trading rules in the appendix in Table~\ref{tab:TradingRules}.

\subsection{Cryptoassets}
\label{sec:Cryptocurrencies}

Cryptoassets are digital assets that utilize cryptographic
primitives and distributed ledger technology (DLT) and represent some economic resource or value to someone. Based on the conceptual design of their underlying transaction settlement layer, they can roughly be divided into cryptoassets that follow Bitcoin’s Unspent Transaction Output (UTXO) model and those that follow Ethereum's account model. Both designs include native tokens, like BTC on the Bitcoin or ETH on the Ethereum~\citep{Buterin2014} ledger. Furthermore, account-model ledgers enable the deployment of arbitrary programs, which are also known as ``smart contracts''. These programs can be used to issue non-native tokens (or simply tokens) representing arbitrary digital and non-digital assets~\citep{auer2023technology}.

Cryptoassets can be mainly seen as speculative virtual assets, which has also been confirmed by a recent empirical user study~\citep{Henry2019}. Regarding the three main functions of money, Bitcoin meets the function of \emph{medium of exchange} but performs poorly as a \emph{unit of account} and as a \emph{store of value}~\citep{Ciaian2016,Yermack2015}. This is because cryptoassets experience high volatility and different trade prices on different exchanges. Further, the most important cryptoasset, Bitcoin, is untethered to other fiat currencies, making its risk mostly impossible to hedge and posing challenges for proper risk management.

The speculative nature of Bitcoin has also been discussed by \cite{Ciaian2016}, who compared Bitcoin to conventional currencies and their primary function as money. Their results stipulate that Bitcoins' attractiveness is the main driver for price formation, followed by market forces as the second driver. On the contrary, macro-financial developments do not determine the Bitcoin price. Therefore, the study concludes that as long as such speculative investments mainly drive the Bitcoin price, no real competition to fiat currency will emerge. We also followed this point of view and conceived Bitcoin as a speculative virtual asset rather than a currency.

It is also well-known that cryptoassets suffer severe limitations: first, there are technical limitations caused by the underlying blockchain technology, which builds the technical foundation of cryptoassets. Public blockchains lack scalability~\citep{gudgeon2019sok} and require relatively high fees for faster transaction confirmation~\citep{Moser2015}. Furthermore, UTXO-ledgers like Bitcoin pose a major ecological problem because of the energy-intensive mining of new coins~\citep{Malone2014}. A recent study conducted by ~\cite{Stoll2019} estimated the power consumption with 45.8 TWh translating into a carbon emission amount ranging from 22.0 to 22.9 Mt CO$_2$, levels between the carbon footprint of Jordan and Sri Lanka.

Second, the overall system design imposes economic limitations as the computational power invested into the network needs to be balanced in equilibrium to avoid the system's collapse. A collapse could happen as soon as the one-off ``stock'' benefit of attacking the network is more attractive than its maintenance as the trust that emerges out of the proof-of-work method is costly and limiting~\citep{Budish2018}.
Also \cite{Ford2019} investigated a potential attack vector that includes an irrational Byzantine attack. The attacker would lose money within the Bitcoin system but has ``hedged'' his bet in a financially connected system like Ethereum, ultimately making an overall profit. The only possible way to circumvent such arbitrary attacks would be to enforce strong identities, contradicting the main design assumptions of primarily honest and rational behavior. A similar contradiction of the necessity to redesign the blockchain system to make it compatible with the real-world legal system is brought forward by \cite{Schuster2019}. He argues that the necessary design changes would classify the claimed advantages of a blockchain irrelevant, hence from his perspective rendering the blockchain as a whole as ``\emph{[...] largely pointless}'' and bringing no benefit to the current economic system. This critical view is shared by \cite{Roubini2019}, who describes the whole crypto economy as a big heist, ``\emph{[...] giving rise to an entire criminal industry, comprising unregulated offshore exchanges, paid propagandists, and an army of scammers looking to fleece retail investors[...]}''.

Despite strong criticism across academic fields, the cryptoasset economy is still a growing business sector. \cite{Hileman2017}, who analyzed nonpublic data for the global cryptoasset market (150 different cryptoassets, covering 38 countries from five world regions), found that the ecosystem is still a rapidly evolving industry. Since Bitcoin was the first cryptoasset in place and is still the most used one with the highest market capitalisation~\citep{Hileman2017}, Bitcoin is the focus of our study.

\subsection{Related Empirical Studies}
\label{sec:EmpiricalEvidence}

The disposition effect is a prominent phenomenon in behavioral finance~\citep{Barberis2013}. Various studies investigating the disposition effect on traditional markets have been conducted, covering different scenarios where the disposition effect seems prevalent. Those studies focused on country-specific markets~\citep{Grinblatt2001,Shapira2001,Shumway2005,Metzger1985,Kaustia2011}, risky asset trading or real estate markets~\citep{Weber1998,Genesove2001}, investor trading behavior~\citep{Zuchel2002}, and market makers~\citep{Coval2005}. The empirical evidence of those studies confirms the existence of the disposition effect. 

Within the cryptoasset area, recent research by \cite{Baur2018} on asymmetric volatility in 20 cryptoasset markets has proposed a link between volatility and disposition effect. The researchers investigated the effect of \emph{fear of missing out} related to uninformed traders in rising market conditions, leading to higher volatility than in falling markets. They argue that the identified asymmetry is in line with the disposition effect but point out that this effect is very weak for two of the biggest cryptoassets, Bitcoin and Ethereum, which are two markets not dominated by uninformed traders. The authors used a model based on (T)GARCH (Generalised Autoregressive Conditional Heteroskedasticity) and quantile regression to estimate volatility. The applied GARCH model sets the primary interest on the (mostly negative) asymmetric volatility indicator $\gamma$. This indicator summarises the trading behavior after positive and negative shocks where uninformed traders are more likely to trade in upward markets and less in downward markets.

In contrast to the study of Baur and Dimpfl, which focuses on the volatility after positive and negative shocks but does not look specifically into the disposition effect, our research tests the hypothesis of the existence of the disposition effect as the primary goal. Our approach focuses on one cryptoasset and measures the frequency of sell activities of investors in upward and downward market conditions, not distinguishing between informed or uninformed traders or positive/negative shock events.

\section{Data and Methods}
\label{sec:DataAndMethod}

The main goal of our approach is to apply the well-known measurement method for the disposition effect on trading activities in Bitcoin. Since buy and sell trades are nowadays executed mostly via (custodial) service providers such as cryptoasset exchanges~\citep{Anderson2018}, we presuppose that transactions from and to exchanges can be regarded as proxies for aggregated, individual trading activities and that they reflect the overall market sentiment.

\subsection{Dataset Collection}

We consider the entire Bitcoin blockchain from its inception until November 24th, 2021 (block 711,189). Since we test our hypothesis on a yearly and monthly basis and therefore consider only completed years. We use the GraphSense Cryptocurrency Analytics Platform~\citep{Haslhofer:2021a} to compute the Bitcoin\emph{entity graph} using the well-known multiple-input address clustering heuristics~\citep{Ron2012,Reid2013a}. The underlying intuition is that if two addresses (e.g., A and B) are used as inputs in the same transaction while one of these addresses along with another address (e.g., B and C) are used as inputs in another transaction, then the three addresses (A, B and C) must somehow be controlled by the same real-world entity~\citep{Meiklejohn2016}, who conducted both transactions and therefore possesses the private keys corresponding to all three addresses. Since this heuristic can fail when CoinJoin transactions~\citep{Moser2017} are involved, we filtered these transactions out using detection heuristics similar to those found in the tool BlockSci~\citep{Kalodner2017} before applying the multiple-input heuristics.

To identify entities representing cryptoasset exchanges, we manually extracted sample addresses of known exchanges from the walletexplorer\footnote{\url{https://www.walletexplorer.com/}} page and mapped them to Bitcoin entities.

\subsection{Dataset Abstraction}

The extracted Bitcoin entity transaction graph represents all Bitcoin transactions between sending and receiving entity, containing all for this research relevant attributes, like the amount of Bitcoins sent, the wallet addresses, and the timestamp when the transaction took place. This data is extracted directly from the publicly available Bitcoin blockchain.

Technically, the resulting data structure is a directed labeled graph in which each node and edge maintains a set of properties. This is also known as property graph~\citep{rodriguez2010constructions}. Assuming that $\boldsymbol{A}$ is a set of addresses, $\boldsymbol{T}$ is the set of transactions in Bitcoin within a certain block range, and $\boldsymbol{C}$ is the set of entity categories known by GraphSense. We can then formalize our model as follows:

\begin{definition}[Entity Graph]\label{def:entity_graph}
	An \emph{entity graph} is a tuple $G = (N,E,\rho,\lambda,\tau,\sigma)$ were:

	\begin{enumerate}
		\item $N$ is a set of nodes representing entities in Bitcoin
		\item $E$ is a set of edges representing transactions between Bitcoin entities
		\item $\rho$ is a function that associates an edge $E$ with a pair of nodes in $N$
		\item $\lambda : N \rightarrow SET^{+}(A)$ is a function that associates a node with a set of addresses from $A$ (i.e., $\lambda$ returns the addresses that are somehow controlled by a certain Bitcoin entity)
		\item $\sigma : N \rightarrow SET^{+}(C)$ is a function that associates a node with a set of categories from $C$. Note that a node can carry several categories (e.g., exchange AND wallet provider)
		\item $\tau : E \rightarrow SET^{+}(T)$ is a function that associates an edge with the set of transactions from $T$, which have taken place between two entities.
	\end{enumerate}

\end{definition}

Given two nodes $n_1, n_2 \in N$ and an edge $e \in E$ such that $\rho(e) = (n_1,n_2)$, we say that $n_1$ and $n_2$ are the \emph{source entity} and the \emph{target entity} of $e$ respectively. Further, we denote $T_{n_1,n_2} \in T$ as the set of transactions that transferred value from a source to a target entity, such that $\tau(e) = T_{n_1,n_2}$.

\begin{figure}
\centering
\includegraphics[width=1\textwidth]{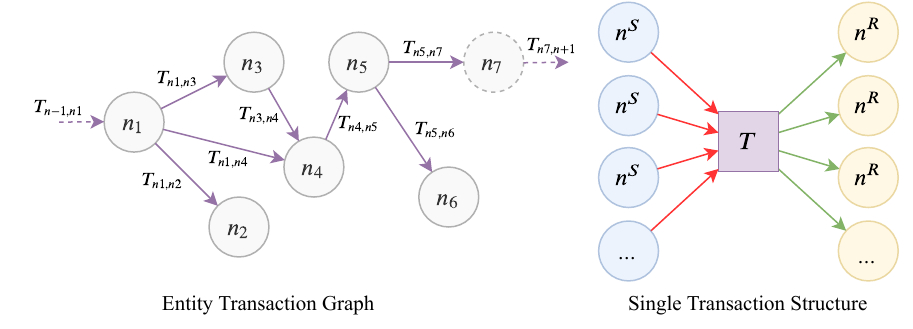}
\caption{Simplified illustrations of an Entity Transaction Graph (left) and the related raw data structure (right) for a simple transaction in the graph, sending entities $n^S$ denoted as input with a \emph{negative} amount, receiving entities $n^R$ denoted as output with a \emph{positive} amount.}
\label{fig:TxInputOutput}
\end{figure}

\subsection{Identifying Selling Transactions}\label{subsec:identifying_sell_txs}

\begin{figure}
	\centering
	\includegraphics[width=1\textwidth]{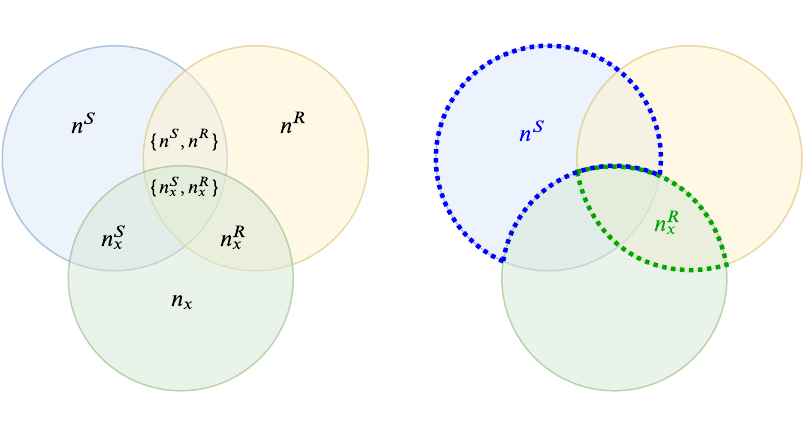}
	\caption{The overall set (left) and the relevant subsets for sending non-exchange $n^S$ and receiving exchange $n^R_x$ entities outlined with dotted lines (right).}
	\label{fig:Venn_01}
\end{figure}

Given the directed nature of the entity graph, an entity can be the sender ($n^S$) or recipient ($n^R$) of transactions. Additionally, some entities in the graph represent cryptoasset exchanges $n_x$ such that $\sigma(n_x) = \{Exchange\}$. Therefore, we denote $n^S_x$ and $n^R_x$ as being sending and receiving exchanges, respectively. Figure~\ref{fig:Venn_01} depicts all relevant subsets required for identifying selling transactions. Simply put, we are extracting the sell transactions where the sending entity is \emph{not} a cryptoasset exchange whereas the receiving entity \emph{is} in fact a cryptoasset exchange entity.

In practice, entities in the entity graph correspond to user wallets or software services that control private keys on behalf of their users. There are two types of wallets, (i) non-custodial, also known as cold or offline wallets and (ii) custodial, hot or online wallets, which are offered by wallet providers~\citep{EcbOnVc2015}. Both types implicate differences in convenience, ease-of-use, and security~\citep{Frantz2016}. Non-custodial wallets are decentralized. The investor owns its private key to access the wallet and has full control but also full responsibility over the funds. This means that if the private key or the restore password gets irrecoverably lost, it is impossible to access the wallet's funds.

Custodial wallets are often integrated into exchange platforms and function similarly to a bank account that stores fiat currency and one or more cryptoassets. A potential investor can initiate a cryptoasset transfer to another custodial or non-custodial wallet as well as initiate a buy or sell transaction on the platform, similarly to exchanging one fiat currency to another. The custodian usually keeps the customer's private keys and also provides backup and accessibility measures for their customers~\citep{EcbOnVc2015}.

Under the assumption that an informed investor holds cryptoasset funds in some non-custodial wallet (e.g., a cold wallet), which has been confirmed by~\cite{Abramova:2021ab}, selling cryptoasset units typically involves several steps: first, the user creates a custodial wallet at some cryptoasset exchange; second, the user transfers funds from his non-custodial wallet to his exchange wallet; and third, the users issues a sell order, which is then executed by the exchange and typically involves the transfer of funds to another custodial wallet, which is assigned to another user, but controlled by the same exchange.

This assumption clearly does not hold for investors who trust the safe custody of their assets in cryptoasset exchanges and keep them in custodial wallets\footnote{Hacks on cryptoassets and other custodial wallet providers have become a major attack vector, with damages exceeding billions of dollar a year~\citep{anderson2019measuring}}. In this case, a sell order changes the balances in the exchange-controlled ledger but does not leave a footprint on the blockchain. Since custodial exchanges are black-boxes, it is currently not possible to reliably assess the extent of such exchange-internal ``off-chain'' transactions~\citep{Anderson2018}. Nevertheless, since our work aims to measure trading behavior under specific market trends rather than quantify overall trading volumes, we assume that we can safely base our further analysis on this assumption.

Therefore, to measure an investor's selling activities, we are interested in the second step within the process mentioned above: the transfer of funds from a non-exchange-controlled wallet to exchange controlled wallets. Hence, we also exclude exchange-to-exchange transactions. More formally, we identify selling transactions by filtering those that have non-exchanges ($n^S$) as source and exchanges as target ($n^R_x$). This specific subset of selling transactions $t_s \subseteq T$ is of main interest in our further analysis.

Table~\ref{tab:YearlyViewExTxMean} represents a high-level, aggregated yearly view of the number of selling transactions and the according transferred (mean) values of those transactions that were received by entities representing exchanges in our dataset. We can clearly see a decline in the number of transactions and the amount of coins (in Bitcoins) transferred from 2017 to 2018, 2019, 2020, and 2021 consecutively. This could be for several reasons: first, the entities on walletexplorer.com might not cover the most recent Bitcoin entity clusters controlled by exchanges; second, it could also reflect that profit-oriented traders and investors (``hodlers'') and rather unexperienced users motivated by fear of missing out (``rookies'') tend to prefer convenient custodial solutions over burdensome non-custodial wallets~\citep{Abramova:2021ab}. In such cases, transactions are not reflected on the blockchain.

\begin{table*}
	\centering
	\resizebox{\textwidth}{!}{
		\begin{tabular}{lrrrrrrrrr}
		\toprule
             & \textbf{2013}      & \textbf{2014}      & \textbf{2015}      & \textbf{2016}      & \textbf{2017}      & \textbf{2018}      & \textbf{2019}      & \textbf{2020} & \textbf{\textbf{2021}}      \\
		\midrule
		No. TxCount      & 1,165,715 &	 2,648,112 &	 3,300,817 &	 5,653,099 &	 9,798,775 &	 3,325,166 &	 2,254,961 &	 2,474,544 &	  1,640,131 \\ \hdashline
		Mean TxValue    & 820.2917 & 940.6239 & 1,854.1545 & 1,585.1754 & 1,262.2556 & 403.1026 & 200.6024 & 86.2578 &  38.2498 \\
		\hdashline
		No. Exchanges & 37        & 99        & 108       & 106       & 97        & 86        & 78        & 70 & 59       \\ \hdashline
		\bottomrule
	\end{tabular}
	}
    \caption{Yearly statistics on the number of sell transactions (No. TxCount) and the related transaction mean value (Mean TxValue) in Bitcoins covered by the identified sell transactions, received by the known exchange entities in our dataset (No. Exchanges). Cut-off date for 2021 was the 24th of November.}
	\label{tab:YearlyViewExTxMean}
\end{table*}

\subsection{Calculating Gains Realized (GR) and Losses Realized (LR)}
\label{subseq:CalcGrLr}

To distinguish if a selling activity should be counted as a gain realized GR or as loss realized LR, we correlate each selling transaction $t_s$ with the market sentiment of the day it has been executed. We define the reference point for this decision as $\bar{\psi}$ for each technical indicator. For the Odean indicator, this is the average of the opening and closing price of the asset in the respective time window, for this study an \emph{hourly} window was used, in contrast to a daily window that is typically the basis for calculating technical indicators. For the technical indicators established buy and sell rules for $\bar{\psi}$ are applied, e.g., MACD buy (GR) for values greater zero, sell (LR) for below zero, similar for RSI buy (GR) for values greater or equal 50 and sell (LR) for values below 50. See appendix Table~\ref{tab:TradingRules} for the complete ruleset similar to a previous study by \cite{Gerritsen2019}.

As a next step, we classified the market sentiment into either upward or downward-facing sentiment. The basis for this was the OHLC (open/high/low/close) data for the Bitcoin market with an hourly resolution, which was acquired from the data broker Kaiko\footnote{Kaiko.com: \url{https://www.kaiko.com/pages/historical-data}} and CryptoSheets\footnote{cryptosheets.com: \url{https://cryptosheets.com/}} via CSV files provided for multiple exchanges, covering the period as early as July 2010 to the 24th November 2021. Like Bitcoinaverage
\footnote{Bitcoinaverage.com: \url{https://bitcoinaverage.com/}}, we used an exchange-independent global average price across all exchanges. Such averaged prices are also used in other empirical research~\citep{Urquhart2016,Spenkelink2014}.

In order to classify Bitcoin selling activities either as GR or LR, we process and refine our dataset as follows: 

\begin{enumerate}

\item OHLC (open, high, low, close) data from the data brokers is joined with the transactions $t_s$ that resulted out of the previous filtering steps

\item For each transaction, we compute $\bar{\psi}$ as the average price (Odean indicator) by averaging the opening and closing price of that hour or $\bar{\psi}$ being the respective reference point for a buy/sell decision for the specific technical indicators

\item For the Odean indicator the market of that day is labeled as either having a \emph{positive} sentiment when the reference point $\bar{\psi}$ is \emph{above} the open price $\psi$ ($\bar{\psi} > \psi$) and \emph{negative} if the average price is \emph{below} the open price $\psi$ ($\bar{\psi} < \psi$)

\item For the technical indicators the market of that day is labeled as either having a \emph{positive} sentiment when the reference point $\bar{\psi}$ indicates a \emph{buy} signal being categorized as GR or having a \emph{negative} sentiment, indicating a \emph{sell} signal being categorized as LR

\item All sell transactions $t_s$ from all identified entities are categorized as either gains realized GR ($t^{UP}_s$) due to positive sentiment, or losses realized LR ($t^{DOWN}_s$) due to negative sentiment and counted per category and per indicator, such that $t_s = t^{UP}_s \cup t^{DOWN}_s$.

\item Finally, for each receiving exchange entity $n \in n_x^R$ within a certain time interval, we count transactions $t^{UP}_s$ with gains realized ($\bar{\psi} > \psi$) and $t^{DOWN}_s$ with losses realized ($\bar{\psi} < \psi$) and sum up the resulting values per indicator.

\end{enumerate}

We implemented our computational method using Python\footnote{Python Programming Language: \url{https://www.python.org/}} and published the method on a Github \url{https://github.com/jschatzmann/CryptoDisposition}.

\section{Analysis and Results}
\label{sec:results}

\emph{Do investors in atypical cryptoasset markets tend to sell their winning positions more frequently than their losing positions?}? If this is the case, the disposition effect is also prevalent in cryptoasset markets, and we can assume that Bitcoin traders act irrational, just like in traditional markets.

\begin{figure}
	\centering
	\includegraphics[width=1\textwidth]{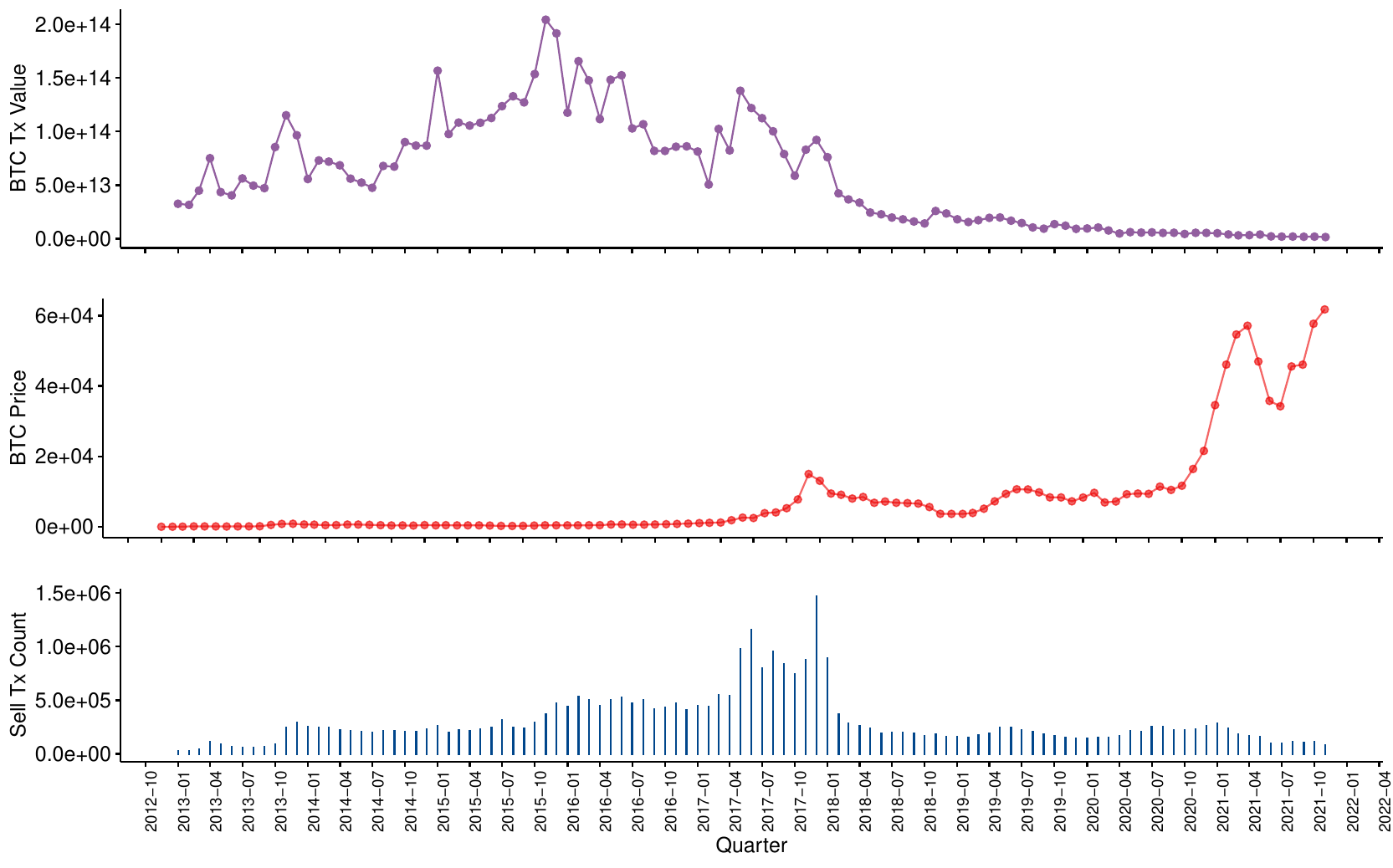}
    \caption{Monthly aggregated amount of sold Bitcoins (\emph{BTC Tx Value} in Satoshis), the Bitcoin price \emph{BTC Price}, and number of identified sell transactions \emph{Sell Tx Count} from 2013 to Nov 2021.}
	\label{fig:ValSumTxCntOverTime}
\end{figure}

Before going into the details of our hypothesis test, we explore, in Figure~\ref{fig:ValSumTxCntOverTime}, how Bitcoin sell transactions evolved with the Bitcoin price. We can observe that before the general Bitcoin publicity boom started in 2016 and took off in 2017, the sold amount of Satoshis peaked in November 2015 with \num{2.04e+14} (2,043,513.89 BTC) at a BTC average price of 356.6 USD. The amount was transferred in a total number of 370,693k sell transactions. In the boom year 2017, the transaction count significantly increased right to the end of December 2017, peaking at 1.47M transactions and \num{9.21e+13} Satoshis (921,237.07 BTC) transferred at a price point of 14.9k USD, flattening out significantly in January 2018, indicating a decrease of investors interests and a tendency to withdraw after the price decline. We also see a steep price increase again in March 2019 and a more extreme price increase starting in July 2020, with consecutive new record highs each month till April 2021. This upward period is followed by a steep price drop in May 2021 while again recovering in August 2021. This drop could be explained by the uncertainty induced by the Covid pandemic, as investors seem to be seeking alternative investment opportunities and are more open to considering cryptoassets in their portfolios. 

\subsection{Exploring Realized Gains and Losses}
\label{subseq:Sumstat}

\begin{figure}
	\centering
    \includegraphics[width=1\textwidth]{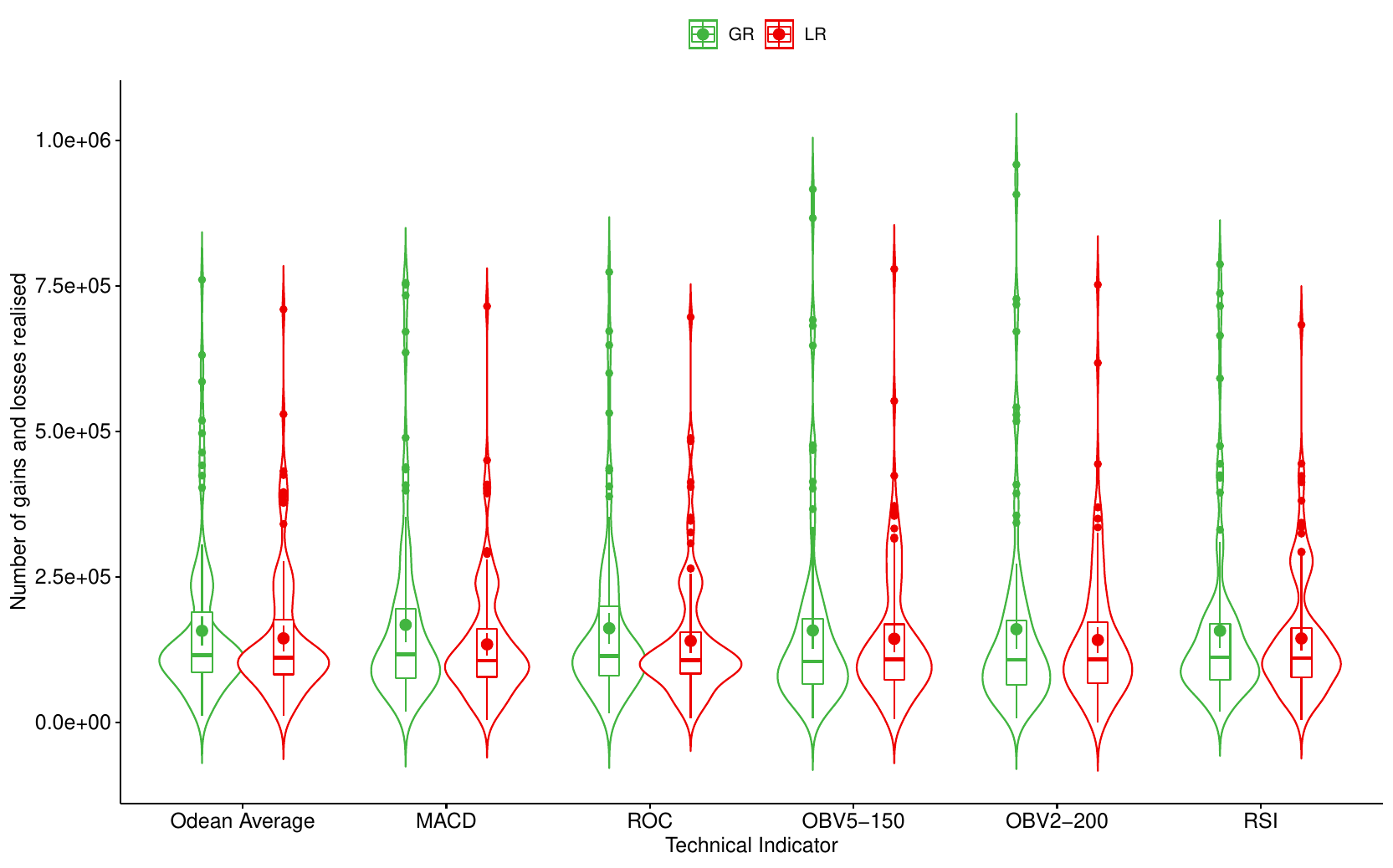}
	\caption{Violin plots of GR and LR for the Odean average, MACD, ROC, OBV5-150, OBV2-200, and RSI, all GR mean values, depicted by the dot marker, are above the LR mean values.}
	\label{fig:ViolinPlotDispEffConf1}
\end{figure}

We computed gains realized (GR), and losses realized (LR) for each indicator and for each entity that can be mapped to a cryptoasset exchange within the Bitcoin ecosystem. In Figure~\ref{fig:ViolinPlotDispEffConf1}, we compare the counts of LR and GR for the first six selected indicators, including the Odean indicator based on the comparison of the closing price with the average (hourly) price. We can observe that the mean values (depicted by the dot marker) of GR lie above the mean values of LR for all technical indicators, providing the basis for the \emph{t}-test. 

\begin{figure}
	\centering
	\includegraphics[width=1\textwidth]{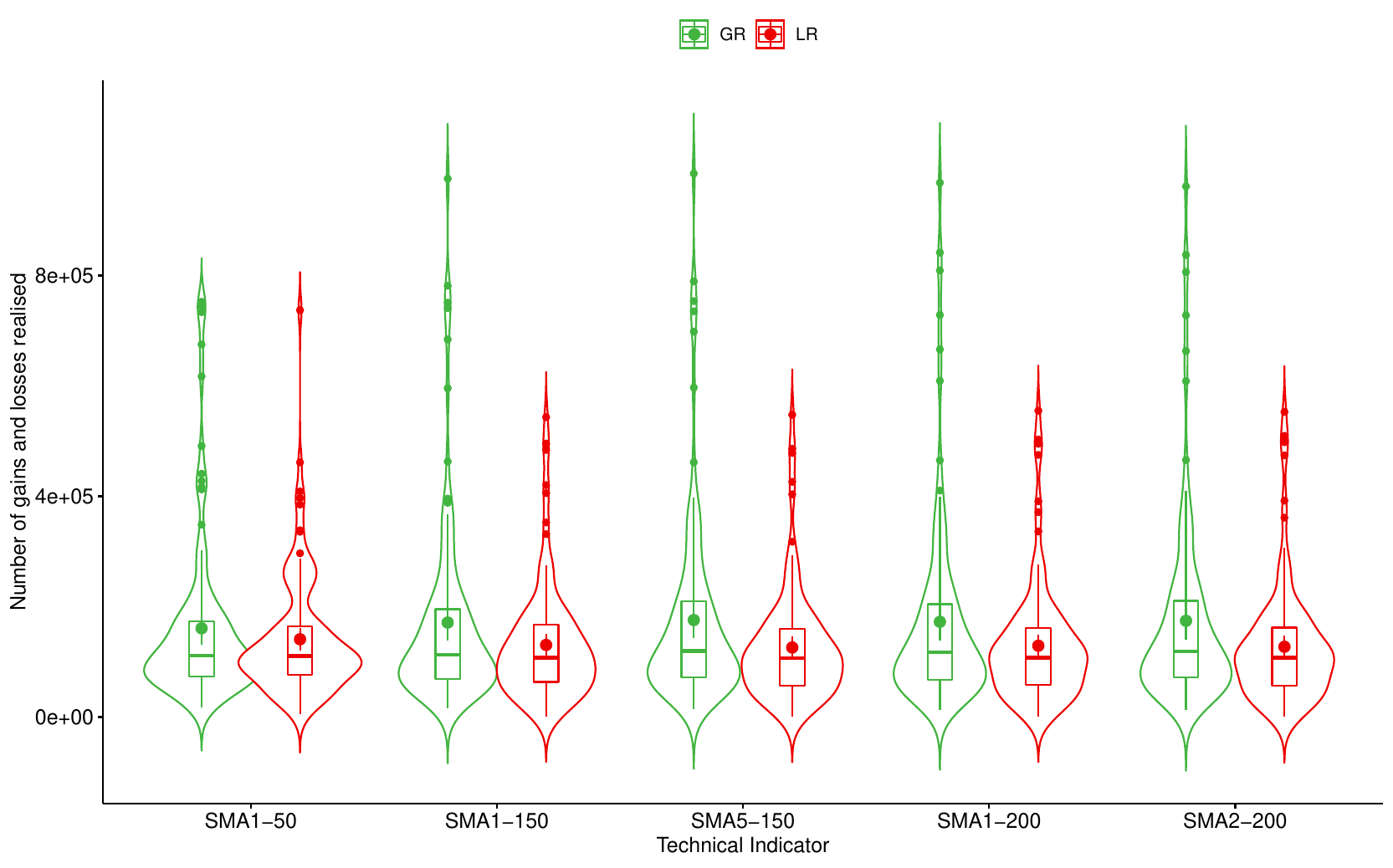}
	\caption{Violin plots for the SMA indicators, the GR mean values, depicted by the dot marker, are all above the LR mean values.}
	\label{fig:ViolinPlotDispEffConf2}
\end{figure}

Similarly, all Simple Moving Average (SMA) indicators (see Appendix~\ref{sec:appendix}) represented in Figure~\ref{fig:ViolinPlotDispEffConf2} over the time window 2013 to 2021 also show that the GR mean values are above LR mean values.

\subsection{Testing for the Disposition Effect in Bitcoin}
\label{subseq:DispEffectInAtypicalMarkets}

We based the measurement method on the model of~\cite{Odean1998} and applied some adaptations to fit the remaining indicators and a portfolio of only one asset (in this case, Bitcoin) and the specific market conditions impacted by the high volatility compared to established markets. These adaptations included removing \emph{relative gains} in the equation, covering the remaining assets in the portfolio and using an adapted calculation method as described in Section~\ref{subseq:CalcGrLr}.

When considering our entire observation period from 2013 to 2021, our empirical result for the Bitcoin market supports the existence of the disposition effect with an overall \textit{t}-statistic of -8.6079 for the Odean average. For seven technical indicators like TRBs, OBV1-50, OBV1-150, OBV1-200, we found an opposite, positive \textit{t}-statistic. Hence, for the indicators Odean, MACD, ROC, all SMAs, OBV5-150, OBV2-200, and RSI, the $H_0$ can be \emph{rejected} with $p \lt 0.001$. For all TRBs, OBV1-50, OBV1-150, OBV2-200, and BB $H_0$ \emph{cannot} be rejected. Table~\ref{tab:t-stat_all_indicators} summarises the results of our analysis aggregated yearly for all indicators in scope. 

\begin{table*}
	\centering
\begin{tabular}{lrrrr}
\toprule
\multicolumn{1}{l}{\textbf{Indicator}} & \multicolumn{1}{r}{\textbf{GR}} & \multicolumn{1}{r}{\textbf{LR}} & \multicolumn{1}{r}{\textbf{tstat}} & \multicolumn{1}{c}{\textbf{pval}} \\
\midrule

Odean average &16,823,282 & 15,435,408 & -8.6097 &\textless{}0.001 \\ \hdashline
MACD &17,921,368 & 14,339,952 & -22.2767 &\textless{}0.001 \\ \hdashline
ROC &17,289,625 & 14,971,524 & -14.3921 &\textless{}0.001 \\ \hdashline
RSI &16,837,629 & 15,423,691 & -8.7712 &\textless{}0.001 \\ \hdashline
SMA 1-50 &17,195,419 & 15,065,901 & -13.2185 &\textless{}0.001 \\ \hdashline
SMA 1-150 &18,304,448 & 13,956,872 & -27.0832 &\textless{}0.001 \\ \hdashline
SMA 5-150 &18,788,227 & 13,473,093 & -33.188 &\textless{}0.001 \\ \hdashline
SMA 1-200 &18,451,757 & 13,809,563 & -28.9376 &\textless{}0.001 \\ \hdashline
SMA 2-200 &18,648,195 & 13,613,125 & -31.4165 &\textless{}0.001 \\ \hdashline
TRB 50 &12,170,633 & 20,090,687 & 49.886 &\textless{}0.001 \\ \hdashline
TRB 150 &12,574,099 & 19,687,221 & 44.6655 &\textless{}0.001 \\ \hdashline
TRB 200 &12,755,731 & 19,505,589 & 42.3305 &\textless{}0.001 \\ \hdashline
OBV 1-50 &8,368,742 & 23,892,578 & 102.551 &\textless{}0.001 \\ \hdashline
OBV 1-150 &8,349,046 & 23,912,274 & 102.8466 &\textless{}0.001 \\ \hdashline
OBV 5-150 &16,897,068 & 15,364,252 & -9.5095 &\textless{}0.001 \\ \hdashline
OBV 1-200 &8,342,269 & 23,919,051 & 102.9483 &\textless{}0.001 \\ \hdashline
OBV 2-200 &17,126,057 & 15,135,263 & -12.3557 &\textless{}0.001 \\ \hdashline
BB &1,199,948 & 2,099,065 & 16.9446 &\textless{}0.001 \\ \hdashline

\bottomrule
\end{tabular}
	\caption{Overview of the \textit{t}-statistic, the number of gains (GR) and losses (LR) realized based on the chosen indicator.}
	\label{tab:t-stat_all_indicators}
\end{table*}

These results show that the original Odean average value as well as MACD, ROC, RSI, all SMAs, and OBV5-150 including OBV2-200 have highly significant negative \textit{t}-values. This result is in line with the findings of \cite{Odean1998}, who utilized trading data of 10,000 accounts with 162,948 records.

A possible explanation for the positive overall \emph{t}-statistic for trading range breakout (TRB) and Boellinger Bands (BB) could be that in the years 2013 to 2016 and even into early 2017 were stable, more linear price changes and no massive, explosive price movement took place. Hence range breakout strategies would not yield excessive gains. Indeed on the yearly view, all TRB indicators indicate statistically significant negative \emph{t}-statistics for the boom and bust year 2017: TRB50 (-29.3079), TRB150 (-40.5594), and TRB200 (-45.0459) having even more extreme values than their SMA counterparts. Taking a closer monthly look into the year 2017 for BB, only January, February, April, May, and November are highly significant ($p \lt 0.001$ ) in the positive range. Only the year 2018 signals a statistically significant negative \emph{t}-statistic of BB (-4.205), assuming the risk-averse investors following a BB strategy tried to mitigate the losses by selling the remaining and newly achieved gains more readily than in the years before. The detailed yearly view is available in the appendix in Table~\ref{tab:TableYearMonth1}, Table~\ref{tab:TableYearMonth2}, and Table~\ref{tab:TableYearMonth3}. The picture for OBV is ambivalent, with three indicators in the plus and two in the minus range.

A possible explanation why TRB, three OBV, and BB indicators do not support our hypothesis could be that TRB is an early indicator for trends that can offer limited downside risk when applied correctly, as well as signaling a starting point for future volatility. Failing or having difficulties adjusting previous support and resistance levels in such a highly volatile market can lead to wrong stop loss limits and hence to higher losses realized. Furthermore, OBV as a momentum indicator that uses volume to predict changes in stock price, likely produces false signals in such hyper volatile markets like Bitcoin. Trading volume spikes within a time frame (hour/day) can push the indicator off for a significant time, making it difficult for investors for making the right call. Using BB as the only indicator is risky as BB only take price and volatility into account, assuming that in case the price deviates substantially from the mean it eventually trends back to this mean. This strategy can fail particularly during strong trends which were seen frequently in the Bitcoin market.

In summary, our study reveals that the disposition effect is prevalent in the Bitcoin market between 2013 and 2021, as indicated by the above statistical tests computed over the entire observation period. Out of the 18 indicators analyzed, 11 demonstrate clear evidence of the disposition effect, including the Odean average indicator.

\subsection{Longitudinal Analysis}\label{sec:results_longitudinal}

\begin{figure*}
\centering
\captionsetup{justification=centering}
\includegraphics[width=1\textwidth]{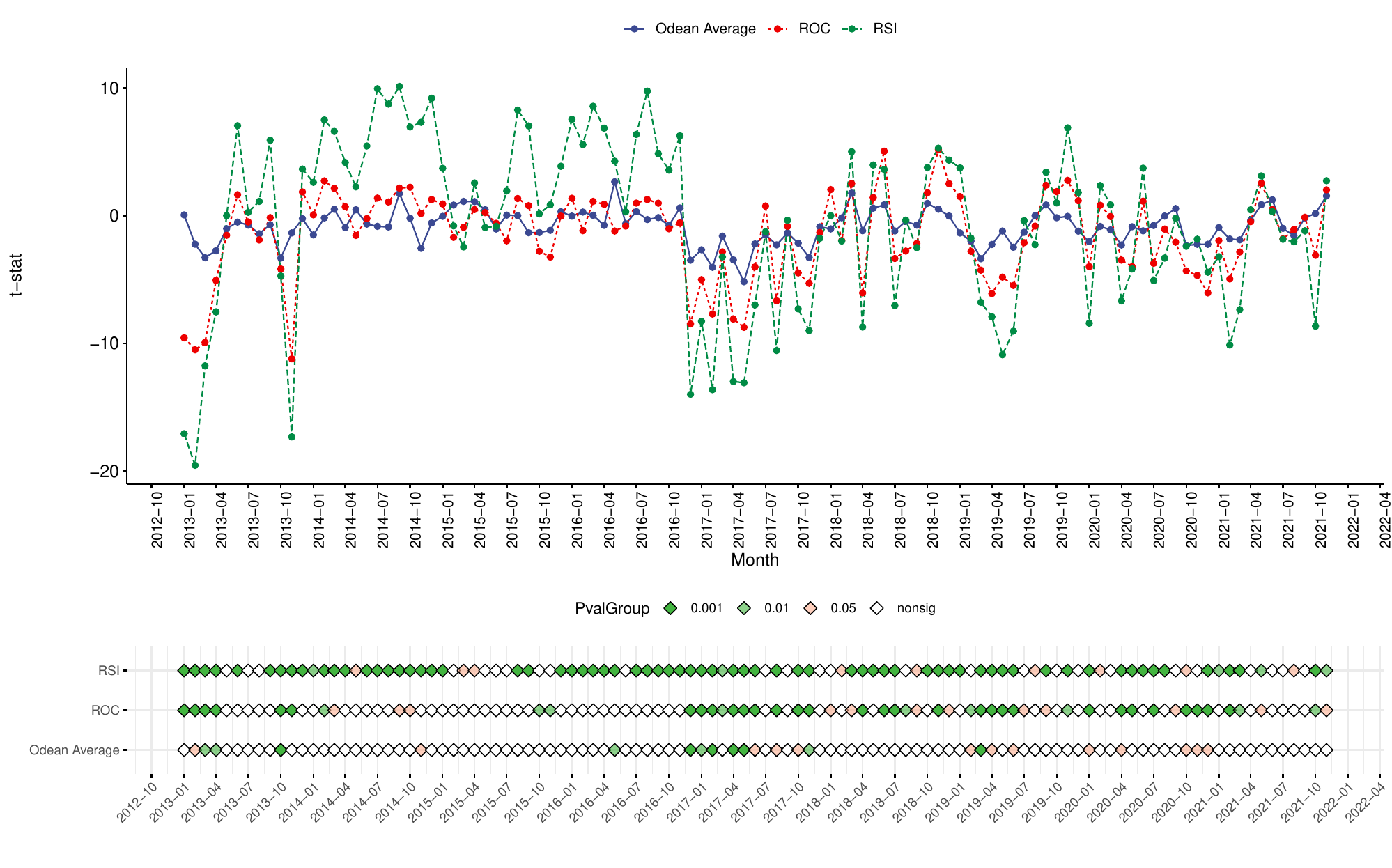}
\caption{Plot of \emph{t}-statistics for Odean, RSI, and ROC GR and LR in a monthly view from Jan-2013 to Nov-2021, including the significance levels.}
\label{fig:TvalPerMonth}
\end{figure*}

In Figure~\ref{fig:TvalPerMonth}, we show the evolution of \textit{t}-statistics over time for selected indicators (Odean, RSI, and ROC) on a monthly basis. The complete results for the entire observation period broken down to monthly statistics is available in Tables~\ref{tab:TableYearMonth1}, \ref{tab:TableYearMonth2}, and \ref{tab:TableYearMonth3} in the Appendix.

When examining monthly RSI values, we can observe significant differences in the \textit{t}-statistic: there are periods in which negative \textit{t}-values indicate strong disposition effect impacted trading (e.g., early 2017 where even TRB indicators signal gains realized) and periods with positive \textit{t}-values supposedly rational trading (e.g., almost the entire years 2014, 2015, and 2016). The \emph{p}-value plot for the three selected indicators in the second row indicates with dark green highly significant $p$-values ($\lt0.001$), light green for $p$-values $\lt0.01$, light red for significant ($p\lt0.05$) values, and white for not significant. A combination of the significance plot and the line plot data signals significant disposition effect-driven trading (negative \emph{t}-statistics) for all three selected indicators. This is true for almost the entire year 2017, changing to a more ambiguous picture in 2018 with mixed signals over the months. 

The lower part in Figure~\ref{fig:TvalPerMonth} indicates the \emph{p}-values for the \emph{t}-statistic given in the upper part of the plot. We see extreme \emph{t}-values in the early years 2013 and 2014 with relatively low sell transaction frequency, low values, and low price, as depicted earlier in Figure~\ref{fig:ValSumTxCntOverTime}. Compared to this, in 2017, where price, sell transactions, and transferred values peaked, we see an almost constant level of significant \emph{t}-values confirming the disposition effect throughout the whole year. Compared to ROC ($n=10$) and RSI ($n=14$), which incorporate $n$ previous time windows for calculating the buy or sell decision, the Odean Average considers only the current time window ($n=1$), which makes it less sensitive for the overall volatility. Hence, the Odean Average (OV) signals disposition effect biased trading only for a few months as depicted in the lower part of Figure~\ref{fig:TvalPerMonth}. Especially the year 2017 has a boosting effect on the overall OV significance.

The major shift in disposition effect impacted trading in 2017 could be explained by changes in the user group around that time: before, technically savvy enthusiasts used and traded Bitcoins mainly. Afterward, casual users, who are curious about the technology and seek long-term financial gains, as well as non-technically savvy traders, joined the market~\citep{Abramova:2021ab}. Since this turning point, most technical indicators signal continuing, stronger disposition effect biased selling activities by the investors.

\section{Discussion}\label{sec:discussion}

\subsection{Key findings}

Our research suggests that the disposition effect is also prevalent among investors in the Bitcoin cryptoasset market. Specifically, we found evidence of this effect across eleven out of eighteen indicators when considering the period from January 2013 to November 2021 on an annual basis.

However, our analysis also shows that the prevalence of the disposition effect in Bitcoin is not consistent on a monthly basis. We found that the effect varies between being significantly negative (confirming the presence of the disposition effect) and being positive (no evidence of the disposition effect) in certain periods. The more stable period stretches from 2013 until the end of 2016 with only limited exceptions of significant negative $t$-statistics for all three selected indicators (Odean, RSI, ROC) in e.g., February, March, April, and October 2013, confirming the disposition effect in those four months. The period from December 2016 till the end of 2017 can be characterized as GR driven, confirming the disposition effect, again followed by a more ambiguous period for 2018 and the following years, yet leaning more towards disposition effect-based trading. The reason for this is yet unclear and would be subject to future research.

Although Bitcoin experiences unusual high volatility, the market itself seems to be controlled by informed investors as found by \cite{Baur2018} and the timing of the sell activities combined with positive or negative shocks require closer examination. Also, the period from January 2017 until December 2017, when fourteen of the indicators signal strong disposition effect require further attention. A possible explanation is that Bitcoin gained much traction in the media, correlating with a steep price increase for the entire year 2017 before coming to an abrupt halt and decline in early 2018, potentially inspiring more non-professional traders to enter the market.

Our hypothesis and key findings are also in line with a recent study by \cite{DiMascioImas2021} who found that professional traders are prone to behavioral biases when buying and selling assets. It shows that also informed traders (e.g., institutional or retail investors and market experts) show skills in buying, yet selling decisions underperform substantially in frequency, substance, and consequence suggesting an asymmetric allocation of cognitive resources.

\subsection{Limitations}

Currently, we quantify the selling activities of clusters by counting sell transactions in different market conditions without considering the amounts of Bitcoins sold. Hence we do not assess the economic impact of a transaction on the investor's bottom line but only the sell activity itself. Despite being a limitation, this is still in line with the approach of \cite{Odean1998}, who also focused on the number of transactions only. The main difference to his approach is that he compared the proportion of gains realized (PGR) and the proportion of losses realized (PLR) based on an investor's portfolio. However, due to the high volatility of Bitcoin and as only one asset was considered, this classification approach yielded no relevant results as the spread of highs and lows was too large and stretched over the average purchase price line, providing no indication of the market situation. Hence, we used the hourly average price $\bar{\psi}$ of open and close price for the Odean indicator and in well-known technical indicators.

Another limitation is that many (uninformed) Bitcoin users nowadays use custodial wallets, which are provided and de-facto controlled by exchanges, instead of running their own clients. Consequently, many trades are executed ``off-chain'' within the shadows of black-box services. Since such transactions are not represented on the blockchain, we cannot identify them via our current approach, and they are currently not considered in our analysis. However, our work presently captures informed traders who transfer their funds to a cryptoasset exchange before executing a sell transaction. Our assumption is backed by a recent study by \cite{HoangBaur2022} stating that investors indeed prefer holding Bitcoin off-exchange in respective private wallets and only transfer them to exchanges in case they intend to sell. The main motivation for storing Bitcoin in non-custodial wallets is risk mitigation. Informed investors tend to place more trust in the blockchain and its wallets than in cryptoasset exchanges, which are considered ``central authorities'' or ``middlemen''.

Finally, our work is limited by the unknown reliability of the Bitcoin co-spend clustering heuristics~\citep{Meiklejohn2016}. Even though this technique has become an integral method in cryptoasset analytics and forensics, it is still difficult to quantify its reliability because comprehensive ground-truth datasets are still missing~\citep{Kappos:2022a}.

\subsection{Future work}

Our research covers the still most important cryptoasset Bitcoin but ignores other altcoins playing an important role in the crypto-economy as well. One could apply our methods and analytics setup for other cryptoassets.

Furthermore, a deeper longitudional investigation would be interesting. Our raw data provides hourly time segments, which allow for an investigation of the selling and buying behavior with finer granularity. This hourly moving time window chosen for this research for the applied indicators yielded further insights compared to the average daily view of the market sentiment. Further analysis of the relationship between the technical indicators themselves and potential correlations and a weighted indicator combining the transaction with the number of Bitcoins sold would be of interest.

Finally, attributing addresses or entities involved in trades would undoubtedly contribute to a better understanding of the factors influencing the evolution of LR and GR. However, collecting so-called \emph{attribution tags} is a resource-intensive data collection process, which commercial tool providers usually implement.

\section{Conclusions}
\label{sec:conclusions}

We investigated the prevalence of the disposition effect in Bitcoin and found that investors in Bitcoin were indeed subject to the disposition effect, with varying intensity. It was prevalent especially from the boom and bust year 2017 onwards, where Bitcoin attracted more investors due to increased publicity. Our findings therefore suggest that irrational investor behavior is also present in atypical markets like Bitcoin. They complement a long line of research in the field of behavioral finance, closing an open gap by confirming the existence of a known economic phenomenon in the still most important cryptoasset market. In addition, we proposed a computational method to quantify the disposition effect, which is tailored to the specific characteristics of cryptoasset transactions and can easily be used for analyzing transactions in other markets if they follow the same transaction model.


\backmatter





\bmhead{Acknowledgments}

This work is partially supported by the Austrian Research Promotion Agency (FFG) under grant COMET K1 Austrian Blockchain Center (ABC) funded within the framework of COMET - Competence Centers for Excellent Technologies by BMK, BMDW and the provinces of Vienna, Lower Austria and Vorarlberg.

\section*{Declarations}


\begin{itemize}
\item Competing interests: The authors have no competing interests to declare that are relevant to the content of this article.
\end{itemize}







\begin{appendices}


\section*{Appendix}\label{sec:appendix}
\begin{landscape}

\begin{table}
    \centering
    \caption{Details of GR and LR on a per year basis 2013 to 2016}

\resizebox{1.3\paperwidth}{!}{
\begin{tabular}{lrrrrrrrrrrrrrrrrr}
    \toprule
                                  & \multicolumn{4}{c}{\textbf{2013}}                          & \multicolumn{4}{c}{\textbf{2014}}                            & \multicolumn{4}{c}{\textbf{2015}}                            & \multicolumn{4}{c}{\textbf{2016}}                            \\
    \midrule
    \multicolumn{1}{c}{Indicator} & GR      & LR        & t-stat   & pval             & GR        & LR        & t-stat   & pval             & GR        & LR        & t-stat   & pval             & GR        & LR        & t-stat   & pval             \\
    \midrule
    
Odean AV  & 613,241 & 552,369   & -3.3323  & \textless{}0.001 & 1,347,114 & 1,300,998 & -1.5404  & 0.1235           & 1,658,945 & 1,641,872 & -0.4277  & 0.6689           & 2,839,138 & 2,813,159 & -0.3985  & 0.6903           \\
\hdashline
MACD      & 728,460 & 437,255   & -16.1673 & \textless{}0.001 & 1,127,514 & 1,520,598 & 13.2593  & \textless{}0.001 & 1,714,743 & 1,586,074 & -3.2250  & 0.0013           & 2,887,270 & 2,765,829 & -1.8631  & 0.0625           \\
\hdashline
ROC       & 669,975 & 495,740   & -9.5818  & \textless{}0.001 & 1,269,776 & 1,378,336 & 3.6284   & \textless{}0.001 & 1,700,683 & 1,600,134 & -2.5196  & 0.0118           & 2,876,776 & 2,776,323 & -1.5410  & 0.1234           \\
\hdashline
RSI       & 675,113 & 490,602   & -10.1534 & \textless{}0.001 & 991,146   & 1,656,966 & 22.8930  & \textless{}0.001 & 1,515,605 & 1,785,212 & 6.7711   & \textless{}0.001 & 2,340,760 & 3,312,339 & 15.0951  & \textless{}0.001 \\
\hdashline
SMA 1-50  & 701,360 & 464,355   & -13.0923 & \textless{}0.001 & 1,025,482 & 1,622,630 & 20.4126  & \textless{}0.001 & 1,601,912 & 1,698,905 & 2.4304   & 0.0151           & 2,487,269 & 3,165,830 & 10.4731  & \textless{}0.001 \\
\hdashline
SMA 1-150 & 788,399 & 377,316   & -23.1691 & \textless{}0.001 & 1,004,870 & 1,643,242 & 21.8964  & \textless{}0.001 & 1,748,481 & 1,552,336 & -4.9201  & \textless{}0.001 & 2,887,084 & 2,766,015 & -1.8574  & 0.0633           \\
\hdashline
SMA 5-150 & 805,816 & 359,899   & -25.2697 & \textless{}0.001 & 1,065,757 & 1,582,355 & 17.5543  & \textless{}0.001 & 1,855,838 & 1,444,979 & -10.3546 & \textless{}0.001 & 3,114,407 & 2,538,692 & -8.8702  & \textless{}0.001 \\
\hdashline
SMA 1-200 & 788,760 & 376,955   & -23.2122 & \textless{}0.001 & 977,077   & 1,671,035 & 23.9223  & \textless{}0.001 & 1,790,192 & 1,510,625 & -7.0226  & \textless{}0.001 & 2,950,429 & 2,702,670 & -3.8034  & \textless{}0.001 \\
\hdashline
SMA 2-200 & 792,371 & 373,344   & -23.6451 & \textless{}0.001 & 1,011,404 & 1,636,708 & 21.4244  & \textless{}0.001 & 1,837,576 & 1,463,241 & -9.4243  & \textless{}0.001 & 3,071,459 & 2,581,640 & -7.5374  & \textless{}0.001 \\
\hdashline
TRB 50    & 552,089 & 613,626   & 3.3687   & \textless{}0.001 & 223,053   & 2,425,059 & 118.9724 & \textless{}0.001 & 115,338   & 3,185,479 & 135.1229 & \textless{}0.001 & 340,119   & 5,312,980 & 131.6352 & \textless{}0.001 \\
\hdashline
TRB 150   & 668,531 & 497,184   & -9.4214  & \textless{}0.001 & 256,848   & 2,391,264 & 110.0666 & \textless{}0.001 & 71,946    & 3,228,871 & 148.0847 & \textless{}0.001 & 329,500   & 5,323,599 & 133.3311 & \textless{}0.001 \\
\hdashline
TRB 200   & 687,012 & 478,703   & -11.4816 & \textless{}0.001 & 268,212   & 2,379,900 & 107.3325 & \textless{}0.001 & 73,945    & 3,226,872 & 147.4302 & \textless{}0.001 & 326,399   & 5,326,700 & 133.8346 & \textless{}0.001 \\
\hdashline
OBV 1-50  & 0       & 1,165,715 & 87.1856  & \textless{}0.001 & 0         & 2,648,112 & 271.1566 & \textless{}0.001 & 0         & 3,300,817 & 176.8783 & \textless{}0.001 & 0         & 5,653,099 & 230.4188 & \textless{}0.001 \\
\hdashline
OBV 1-150 & 0       & 1,165,715 & 87.1856  & \textless{}0.001 & 0         & 2,648,112 & 271.1566 & \textless{}0.001 & 0         & 3,300,817 & 176.8783 & \textless{}0.001 & 0         & 5,653,099 & 230.4188 & \textless{}0.001 \\
\hdashline
OBV 5-150 & 671,385 & 494,330   & -9.7386  & \textless{}0.001 & 986,323   & 1,661,789 & 23.2450  & \textless{}0.001 & 1,894,062 & 1,406,755 & -12.3119 & \textless{}0.001 & 2,440,428 & 3,212,671 & 11.9411  & \textless{}0.001 \\
\hdashline
OBV 1-200 & 0       & 1,165,715 & 87.1856  & \textless{}0.001 & 0         & 2,648,112 & 271.1566 & \textless{}0.001 & 0         & 3,300,817 & 176.8783 & \textless{}0.001 & 0         & 5,653,099 & 230.4188 & \textless{}0.001 \\
\hdashline
OBV 2-200 & 672,718 & 492,997   & -9.8868  & \textless{}0.001 & 958,306   & 1,689,806 & 25.3086  & \textless{}0.001 & 1,919,793 & 1,381,024 & -13.6384 & \textless{}0.001 & 2,547,124 & 3,105,975 & 8.6081   & \textless{}0.001 \\
\hdashline
BB        & 35,331  & 84,595    & 7.7159   & \textless{}0.001 & 19,659    & 145,629   & 17.5810  & \textless{}0.001 & 43,001    & 164,295   & 12.0020  & \textless{}0.001 & 39,963    & 350,591   & 18.7162  & \textless{}0.001 \\

\bottomrule
\end{tabular}
}
    
    \label{tab:TableYearMonth1}

    \caption{Details of GR and LR on a per year basis 2017 to 2020}

\resizebox{1.3\paperwidth}{!}{
\begin{tabular}{lrrrrrrrrrrrrrrrr}
    \toprule
    \multicolumn{1}{c}{}          & \multicolumn{4}{c}{\textbf{2017}}                            & \multicolumn{4}{c}{\textbf{2018}}                            & \multicolumn{4}{c}{\textbf{2019}}     & \multicolumn{4}{c}{\textbf{2020}}                             \\
    \midrule
    \multicolumn{1}{c}{Indicator} & GR        & LR        & t-stat   & pval             & GR        & LR        & t-stat   & pval             & GR        & LR        & t-stat    & pval  & GR        & LR        & t-stat    & pval             \\
    \midrule

Odean AV  & 5,365,974 & 4,431,078 & -7.7887  & \textless{}0.001 & 1,677,114 & 1,648,052 & -0.5938  & 0.5527           & 1,184,141 & 1,070,820 & -4.1929   & \textless{}0.001 & 1,298,375 & 1,176,169 & -4.2168   & \textless{}0.001 \\
\hdashline
MACD      & 6,225,831 & 3,572,944 & -22.6627 & \textless{}0.001 & 1,625,319 & 1,699,847 & 1.5230   & 0.1278           & 1,215,714 & 1,039,247 & -6.5388   & \textless{}0.001 & 1,460,088 & 1,014,456 & -15.5723  & \textless{}0.001 \\
\hdashline
ROC       & 5,698,906 & 4,099,869 & -13.4111 & \textless{}0.001 & 1,628,453 & 1,696,713 & 1.3949   & 0.1631           & 1,205,961 & 1,049,000 & -5.8130   & \textless{}0.001 & 1,372,394 & 1,102,150 & -9.3619   & \textless{}0.001 \\
\hdashline
RSI       & 6,147,085 & 3,651,690 & -21.2454 & \textless{}0.001 & 1,652,344 & 1,672,822 & 0.4184   & 0.6757           & 1,236,571 & 1,018,390 & -8.0949   & \textless{}0.001 & 1,357,405 & 1,117,139 & -8.3147   & \textless{}0.001 \\
\hdashline
SMA 1-50  & 6,214,823 & 3,583,952 & -22.4637 & \textless{}0.001 & 1,613,319 & 1,711,847 & 2.0136   & 0.0441           & 1,217,403 & 1,037,558 & -6.6645   & \textless{}0.001 & 1,406,030 & 1,068,514 & -11.7253  & \textless{}0.001 \\
\hdashline
SMA 1-150 & 6,644,609 & 3,154,166 & -30.4806 & \textless{}0.001 & 1,484,339 & 1,840,827 & 7.3060   & \textless{}0.001 & 1,218,938 & 1,036,023 & -6.7789   & \textless{}0.001 & 1,545,705 & 928,839   & -21.8345  & \textless{}0.001 \\
\hdashline
SMA 5-150 & 6,681,329 & 3,117,446 & -31.1926 & \textless{}0.001 & 1,488,986 & 1,836,180 & 7.1144   & \textless{}0.001 & 1,216,021 & 1,038,940 & -6.5616   & \textless{}0.001 & 1,575,797 & 898,747   & -24.0993  & \textless{}0.001 \\
\hdashline
SMA 1-200 & 6,731,901 & 3,066,874 & -32.1813 & \textless{}0.001 & 1,427,330 & 1,897,836 & 9.6647   & \textless{}0.001 & 1,215,164 & 1,039,797 & -6.4978   & \textless{}0.001 & 1,588,534 & 886,010   & -25.0698  & \textless{}0.001 \\
\hdashline
SMA 2-200 & 6,716,247 & 3,082,528 & -31.8742 & \textless{}0.001 & 1,429,305 & 1,895,861 & 9.5827   & \textless{}0.001 & 1,213,859 & 1,041,102 & -6.4006   & \textless{}0.001 & 1,596,026 & 878,518   & -25.6442  & \textless{}0.001 \\
\hdashline
TRB 50    & 6,583,589 & 3,215,186 & -29.3079 & \textless{}0.001 & 1,707,353 & 1,617,813 & -1.8298  & 0.0673           & 1,269,507 & 985,454   & -10.5664  & \textless{}0.001 & 371,705   & 2,102,839 & 77.4491   & \textless{}0.001 \\
\hdashline
TRB 150   & 7,140,103 & 2,658,672 & -40.5594 & \textless{}0.001 & 1,486,842 & 1,838,324 & 7.2028   & \textless{}0.001 & 1,259,840 & 995,121   & -9.8389   & \textless{}0.001 & 323,146   & 2,151,398 & 85.2279   & \textless{}0.001 \\
\hdashline
TRB 200   & 7,343,124 & 2,455,651 & -45.0459 & \textless{}0.001 & 1,453,758 & 1,871,408 & 8.5693   & \textless{}0.001 & 1,256,355 & 998,606   & -9.5771   & \textless{}0.001 & 323,269   & 2,151,275 & 85.2069   & \textless{}0.001 \\
\hdashline
OBV 1-50  & 0         & 9,798,775 & 165.4584 & \textless{}0.001 & 1,999,106 & 1,326,060 & -13.9030 & \textless{}0.001 & 2,254,961 & 0         & -183.8422 & \textless{}0.001 & 2,474,544 & 0         & -207.0634 & \textless{}0.001 \\
\hdashline
OBV 1-150 & 0         & 9,798,775 & 165.4584 & \textless{}0.001 & 1,979,410 & 1,345,756 & -13.0729 & \textless{}0.001 & 2,254,961 & 0         & -183.8422 & \textless{}0.001 & 2,474,544 & 0         & -207.0634 & \textless{}0.001 \\
\hdashline
OBV 5-150 & 6,444,940 & 3,353,835 & -26.6886 & \textless{}0.001 & 1,296,874 & 2,028,292 & 15.1391  & \textless{}0.001 & 1,099,149 & 1,155,812 & 2.0950    & 0.0362           & 1,231,592 & 1,242,952 & 0.3916    & 0.6954           \\
\hdashline
OBV 1-200 & 0         & 9,798,775 & 165.4584 & \textless{}0.001 & 1,972,633 & 1,352,533 & -12.7879 & \textless{}0.001 & 2,254,961 & 0         & -183.8422 & \textless{}0.001 & 2,474,544 & 0         & -207.0634 & \textless{}0.001 \\
\hdashline
OBV 2-200 & 6,712,002 & 3,086,773 & -31.7911 & \textless{}0.001 & 1,211,602 & 2,113,564 & 18.7978  & \textless{}0.001 & 1,061,410 & 1,193,551 & 4.8911    & \textless{}0.001 & 1,207,620 & 1,266,924 & 2.0447    & 0.0409           \\
\hdashline
BB        & 489,363   & 751,446   & 6.1116   & \textless{}0.001 & 229,875   & 164,821   & -4.2050  & \textless{}0.001 & 149,253   & 157,529   & 0.8136    & 0.4159           & 92,868    & 169,354   & 8.1111    & \textless{}0.001 \\

\bottomrule
\end{tabular}
}
    
    \label{tab:TableYearMonth2}
\end{table}
\end{landscape}

\begin{table}
    \centering
    \caption{Details of GR and LR on a per year basis for 2021}

\resizebox{0.5\paperwidth}{!}{
\begin{tabular}{lrrrrrrrrrrrrrrrr}
    \toprule
    \multicolumn{1}{c}{}          & \multicolumn{4}{c}{\textbf{2021}} \\
    \midrule
    \multicolumn{1}{c}{Indicator} & GR        & LR        & t-stat   & pval \\
    \midrule

Odean AV  & 839,240   & 800,891 & -1.7859   & 0.0742           \\
\hdashline
MACD      & 936,429   & 703,702 & -10.9181  & \textless{}0.001 \\
\hdashline
ROC       & 866,701   & 773,259 & -4.3561   & \textless{}0.001 \\
\hdashline
RSI       & 921,600   & 718,531 & -9.5095   & \textless{}0.001 \\
\hdashline
SMA 1-50  & 927,821   & 712,310 & -10.0995  & \textless{}0.001 \\
\hdashline
SMA 1-150 & 982,023   & 658,108 & -15.3058  & \textless{}0.001 \\
\hdashline
SMA 5-150 & 984,276   & 655,855 & -15.5252  & \textless{}0.001 \\
\hdashline
SMA 1-200 & 982,370   & 657,761 & -15.3395  & \textless{}0.001 \\
\hdashline
SMA 2-200 & 979,948   & 660,183 & -15.1039  & \textless{}0.001 \\
\hdashline
TRB 50    & 1,007,880 & 632,251 & -17.8417  & \textless{}0.001 \\
\hdashline
TRB 150   & 1,037,343 & 602,788 & -20.7840  & \textless{}0.001 \\
\hdashline
TRB 200   & 1,023,657 & 616,474 & -19.4096  & \textless{}0.001 \\
\hdashline
OBV 1-50  & 1,640,131 & 0       & -151.2493 & \textless{}0.001 \\
\hdashline
OBV 1-150 & 1,640,131 & 0       & -151.2493 & \textless{}0.001 \\
\hdashline
OBV 5-150 & 832,315   & 807,816 & -1.1408   & 0.2540           \\
\hdashline
OBV 1-200 & 1,640,131 & 0       & -151.2493 & \textless{}0.001 \\
\hdashline
OBV 2-200 & 835,482   & 804,649 & -1.4358   & 0.1511           \\
\hdashline
BB        & 100,635   & 110,805 & 1.2905    & 0.1969          \\

\bottomrule
\end{tabular}
}
    
    \label{tab:TableYearMonth3}
\end{table}

\clearpage

\begin{table*}
    \centering
    \caption{Applied trading rules to identify GR and LR}
    
\resizebox{0.8\paperwidth}{!}{
\begin{tabular}{clclcc}
\toprule
\textbf{Indicator}               & \textbf{Reference 1} & \textbf{Operator} & \textbf{Reference 2} & \textbf{Decision} & \multicolumn{1}{l}{\textbf{Description (time units)}}            \\
\midrule
\multirow{2}{*}{Odean indicator} & Average Price        & \textgreater{}    & Open Price           & GR                & \multirow{2}{*}{\makecell{Custom indicator \\ (1 unit)}}                       \\
                                 & Average Price        & \textless{}       & Open Price           & LR                &                                                                  \\
\hdashline
\multirow{2}{*}{SMA1-50}         & Close Price          & \textgreater{}    & SMA50                & GR                & \multirow{2}{*}{\makecell{Simple Moving Average \\ (short 1, long 50)}}        \\
                                 & Close Price          & \textless{}       & SMA50                & LR                &                                                                  \\
\hdashline
\multirow{2}{*}{SMA1-150}        & Close Price          & \textgreater{}    & SMA150               & GR                & \multirow{2}{*}{\makecell{Simple Moving Average \\ (short 1, long 150)}}       \\
                                 & Close Price          & \textless{}       & SMA150               & LR                &                                                                  \\
\hdashline
\multirow{2}{*}{SMA5-50}         & SMA5                 & \textgreater{}    & SMA5-150             & GR                & \multirow{2}{*}{\makecell{Simple Moving Average \\ (short 5, long 50)}}        \\
                                 & SMA5                 & \textless{}       & SMA5-150             & LR                &                                                                  \\
\hdashline
\multirow{2}{*}{SMA1-200}        & Close Price          & \textgreater{}    & SMA200               & GR                & \multirow{2}{*}{\makecell{Simple Moving Average \\ (short 1, long 200)}}       \\
                                 & Close Price          & \textless{}       & SMA200               & LR                &                                                                  \\
\hdashline
\multirow{2}{*}{SMA2-200}        & SMA2                 & \textgreater{}    & SMA200               & GR                & \multirow{2}{*}{\makecell{Simple Moving Average \\ (short 2, long 200)}}       \\
                                 & SMA2                 & \textless{}       & SMA200               & LR                &                                                                  \\
\hdashline
\multirow{2}{*}{TRB50}           & Close Price          & \textgreater{}    & TRB50 mband          & GR                & \multirow{2}{*}{\makecell{Trading Range Breakout \\ (50 units)}}               \\
                                 & Close Price          & \textless{}       & TRB50 mband          & LR                &                                                                  \\
\hdashline
\multirow{2}{*}{TRB150}          & Close Price          & \textgreater{}    & TRB150 mband         & GR                & \multirow{2}{*}{\makecell{Trading Range Breakout \\ (150 units)}}              \\
                                 & Close Price          & \textless{}       & TRB150 mband         & LR                &                                                                  \\
\hdashline
\multirow{2}{*}{TRB200}          & Close Price          & \textgreater{}    & TRB200 mband         & GR                & \multirow{2}{*}{\makecell{Trading Range Breakout \\ (200 units)}}              \\
                                 & Close Price          & \textless{}       & TRB200 mband         & LR                &                                                                  \\
\hdashline
\multirow{2}{*}{MACD}            & MACD                 & \textgreater{}    & Zero                 & GR                & \multirow{2}{*}{\makecell{Moving Average Convergence \\ Divergence (9 units)}} \\
                                 & MACD                 & \textless{}       & Zero                 & LR                &                                                                  \\
\hdashline
\multirow{2}{*}{ROC}             & ROC                  & \textgreater{}    & Zero                 & GR                & \multirow{2}{*}{\makecell{Rate Of Change \\ (10 units)}}                       \\
                                 & ROC                  & \textless{}       & Zero                 & LR                &                                                                  \\
\hdashline
\multirow{2}{*}{OBV1-50}         & Close Price          & \textgreater{}    & OBV SMA50            & GR                & \multirow{2}{*}{\makecell{On Balance Volume \\ (short 1, long 50)}}            \\
                                 & Close Price          & \textless{}       & OBV SMA50            & LR                &                                                                  \\
\hdashline
\multirow{2}{*}{OBV1-150}        & Close Price          & \textgreater{}    & OBV SMA150           & GR                & \multirow{2}{*}{\makecell{On Balance Volume \\ (short 1, long 150)}}           \\
                                 & Close Price          & \textless{}       & OBV SMA150           & LR                &                                                                  \\
\hdashline
\multirow{2}{*}{OBV5-150}        & OBV SMA5             & \textgreater{}    & OBV SMA150           & GR                & \multirow{2}{*}{\makecell{On Balance Volume \\ (short 5, long 150)}}           \\
                                 & OBV SMA5             & \textless{}       & OBV SMA150           & LR                &                                                                  \\
\hdashline
\multirow{2}{*}{OBV1-200}        & Close Price          & \textgreater{}    & OBV SMA200           & GR                & \multirow{2}{*}{\makecell{On Balance Volume \\ (short 1, long 200)}}           \\
                                 & Close Price          & \textless{}       & OBV SMA200           & LR                &                                                                  \\
\hdashline
\multirow{2}{*}{OBV2-200}        & OBV SMA2             & \textgreater{}    & OBV SMA200           & GR                & \multirow{2}{*}{\makecell{On Balance Volume \\ (short 2, long 200)}}           \\
                                 & OBV SMA2             & \textless{}       & OBV SMA200           & LR                &                                                                  \\
\hdashline
\multirow{2}{*}{RSI}             & RSI                  & \textgreater{}=   & 50                   & GR                & \multirow{2}{*}{\makecell{Relative Strenght \\ Indicator (14 units)}}          \\
                                 & RSI                  & \textless{}       & 50                   & LR                &                                                                  \\
\hdashline
\multirow{3}{*}{BB}              & Close Price          & \textless{}       & BB low               & GR                & \multirow{3}{*}{\makecell{Boellinger Bands \\ (20 units)}}                     \\
                                 & Close Price          & \textgreater{}    & BB high              & LR                &                                                                  \\
                                 & {[}Otherwise{]}      & =                 & {[}Neutral{]}        & N                 &                                                                 \\
\bottomrule
\end{tabular}
}
    
    \label{tab:TradingRules}
\end{table*}

\section*{Technical Indicator Definition}

The following short descriptions and formulas are based on definitions given in~\citep{Kirkpatrick2007,Achelis2003} and~\citep{Murphy1999} as well as the technical analysis library~\footnote{\url{https://www.ta-lib.org/}, \url{https://tulipindicators.org/}} underlying the used Python TA-lib.

\paragraph{SMA - Simple Moving Average}
An SMA shows the average price of an asset or security over a specified period. It is a commonly used smoothing function on time series data. The main parameter $n$ defines the time window for the calculation. The SMA applies equal weight on each price compared to exponential, triangular, or variable moving averages using different weights.

\begin{equation}
sma_{t} = \frac{1}{n} \sum_{i=0}^{n-1} in_{t-i}
\end{equation}

\paragraph{TRB - Trading Range Breakout (Donchian Channel)}
Trading Range Breakout systems generate buy and sell signals when the price moves out of the channel band, depending on the $n$ number of periods for the calculation. The goal of this indicator is to identify bullish and bearish extremes, the middle band is the average of the highest high and the lowest low for $n$ periods.
\begin{align}
\begin{split}
MC&=\frac{UC-LC}{2} \\
  \text{where}~UC &= \text{Highest High in last $n$ periods (upper channel),} \\
  LC &= \text{Lowest Low in Last $n$ periods (lower channel),} \\
  MC &= \text{middle channel,} \\
  n &= \text{number of minutes, hours, days, weeks, months,} \\
  periods &= \text{minutes, hours, days, weeks, months}
\end{split}
\end{align}

\paragraph{MACD - Moving Average Convergence Divergence}
The MACD indicator helps follow trends and takes three parameters: a short period $n$, a long period $m$, and a signal period $p$. It is calculated by subtracting the short from the long period resulting in a value oscillating above and below zero, signaling market trend (above zero bullish, below zero bearish).

\begin{align}
\begin{split}
short_{t} &= ema(n, input) \\
long_{t} &= ema(m, input) \\
macd_{t} &= short_{t} - long_{t} \\
signal_{t} &= ema(p, macd_{t}) \\
histogram_{t} &= macd_{t} - signal_{t}
\end{split}
\end{align}

\paragraph{EMA - Exponential Moving Average}
The EMA applies an exponential smoothing function. It puts greater weight on the significance of the more recent price values and takes one parameter, the period $n$. Larger values for $n$ will result in higher smoothing effects while also creating more lag. The initial calculation value is setting the first EMA output to the first input. The relation for $0 < a \leq 1$ will be satisfied in the first step.

\begin{align}
\begin{split}
a & = \frac{2}{n+1} \\
ema_{t} &= (1-a)ema_{t-1} + (a)in_{t} \\
\end{split}
\end{align}

\paragraph{ROC - Rate of Change}
The ROC indicator calculates the change between the current price and the price $n$ bars ago, providing information on momentum as the speed of e.g. price change. It takes also one parameter for the period $n$.
\begin{equation}
roc_{t} = \frac{in_{t}-in_{t-n}}{in_{t-n}}
\end{equation}

\paragraph{OBV - On Balance Volume}
The OBV indicator calculates the running total of volume by summing up in up-days and subtracted on down-days. It is a momentum indicator providing crowd sentiment information on potential upcoming price changes using trading volume.

\begin{equation}
obv_{t} =\begin{cases} obv_{t-1} + volume_{t} & \mathrm{if} \;  close_{t} > close_{t-1} \\
obv_{t-1} - volume_{t} & \mathrm{if} \;  close_{t} < close_{t-1} \\
0 & \text{else} \end{cases}
\end{equation}

\paragraph{RSI - Relative Strength Indicator}
The RSI is a momentum oscillator. It helps to identify bullish or bearish trends, using one parameter, the period $n$. A asset is assumed overbought when RSI is above 70\% and oversold when below 30\%.
\begin{align}
\begin{split}
up_{t} &=
    \begin{cases}
    in_{t} - in_{t-1} & \mathrm{if} \; in_{t} > in_{t-1} \\
    0 & \mathrm{else}
    \end{cases} \\
down_{t} &=
    \begin{cases}
    in_{t-1} - in_{t} & \mathrm{if} \; in_{t} < in_{t-1} \\
    0 & \mathrm{else}
    \end{cases} \\
sup_{t} &= \frac{n-1}{n} sup_{t-1} + \frac{1}{n} up_{t} \\
sdown_{t} &= \frac{n-1}{n} sdown_{t-1} + \frac{1}{n} down_{t} \\
rsi_{t} &= 100 - \frac{100}{1 + \frac{sup_{t}}{sdown_{t}}}
\end{split}
\end{align}

\paragraph{BB - Boellinger Bands}
The BB indicator calculates a middle band (Simple Moving Average), as well as upper and lower bands. Those bands have an offset of the middle band. BB takes two parameters, the period $n$ and a scaling value $a$. The offset of the upper and lower bands of the middle band is defined by standard deviations of the input value.

\begin{align}
\begin{split}
bbands^{middle}_{t} &= \frac{1}{n} \sum_{i=0}^{n-1} in_{t-i} \\
bbands^{lower}_{t} &= bbands^{middle}_{t} - a \sqrt {\frac{1}{n} \sum_{i=0}^{n-1} (in_{t-i}-bbands^{middle}_{t})^{2}} \\
bands^{upper}_{t} &= bbands^{middle}_{t} + a \sqrt {\frac{1}{n} \sum_{i=0}^{n-1} (in_{t-i}-bbands^{middle}_{t})^{2}} \\
\end{split}
\end{align}







\end{appendices}


\bibliography{main.bib}

\begin{thebibliography}{}
\bibitem [\protect \citeauthoryear {%
van Wijk%
}{%
van Wijk%
}{%
{\protect \APACyear {2013}}%
}]{%
VanWijk2013}
\APACinsertmetastar {%
VanWijk2013}%
\begin{APACrefauthors}%
van Wijk, D.%
\end{APACrefauthors}%
\unskip\
\newblock
\APACrefYearMonthDay{2013}{}{}.
\newblock
{\BBOQ}\APACrefatitle {What can be expected from the BitCoin} {What can be expected from the bitcoin}{\BBCQ}\ [Thesis, ID 345986].
\newblock
\APACjournalVolNumPages{Erasmus Universiteit Rotterdam}{}{}{}.
\newblock

\newblock

\PrintBackRefs{\CurrentBib}
\end{thebibliography}


\end{document}